\documentclass[english,twocolumn,aps,prd,longbibliography,crop,tikz,superscriptaddress]{revtex4-1}
\usepackage[T1]{fontenc}
\usepackage{graphicx}
\usepackage{babel}
\usepackage{braket}
\usepackage{qcircuit}
\usepackage{xcolor}
\usepackage{amsmath}
\usepackage{lipsum}
\usepackage{blindtext}
\usepackage{upgreek}

\usepackage{qcircuit}

\usepackage[colorlinks,linkcolor=black,citecolor=black,urlcolor=black,bookmarks=false,hypertexnames=true]{hyperref} 

\usepackage{fancyhdr}

\fancypagestyle{fancyplain}
\begin{document}

\title{Ground-state energies of Ising models calculated using the samples from a quantum computer that simulates short-time evolution}

\author{John P. T. Stenger}
\thanks{Corresponding Author: john.p.stenger2.civ@nrl.navy.mil}
\affiliation{U.S. Naval Research Laboratory, Washington, DC 20375, United States}
\author{C. Stephen Hellberg}
\affiliation{U.S. Naval Research Laboratory, Washington, DC 20375, United States}
\author{Daniel Gunlycke}
\affiliation{U.S. Naval Research Laboratory, Washington, DC 20375, United States}

\begin{abstract}
We find the ground-state energy of the Ising model using the Cascaded Variational Quantum Eigensolver (CVQE) algorithm with the Guided-Sampling Ansatz (GSA)  using up to 63 qubits on a quantum computer.  We study a heavy-hex lattice to match the qubit architecture, allowing us to perform calculations in the quantum utility regime.  We study both a homogeneous and random-coupling model. We locate the boundary of acceptable quantum errors as a function of the number of qubits and coupling strength.  An entropic analysis is performed giving insights into the quantum computing performance.  A subspace analysis is performed that suggests that the Ising model is especially suited for near-term quantum computing.     
\end{abstract}

\maketitle
\thispagestyle{fancyplain}

\section{Introduction}

Quantum computing promises solutions to certain problems that are intractable on classical computers~\cite{Feynman1982}.  However, many of the quantum algorithms that solve these problems require quantum computers with long coherence times~\cite{kitaev1995,Farhi2001,smelyanskiy2002,Reichardt2004}.  Such algorithms are not computable on near-term quantum devices that suffer from many sources of error.  However, it is beginning to become clear that there is a subset of problems that are intractable on classical computers but may be solvable on quantum computers with short coherence times~\cite{Peruzzo2014,OMalley2016,McArdle2020,jamet2021,Baker2024}.  The solutions to these problems rely on the development of quantum algorithms with shallow circuits.  Many such algorithms are currently being explored, such as the Variation Quantum Eigensolver (VQE)~\cite{Peruzzo2014,McClean2016,OMalley2016,Kandala2017,Wang2019,McArdle2020,Arute2020,Gonthier2020,HeadMarsden2021,Cerezo2021}, the Cascaded Variational Quantum Eigensolver (CVQE)~\cite{Gunlycke2024,Stenger2024,gunlycke2025}, Krylov Eigensolvers~\cite{parrish2019,Stair2020,Baker2021,jamet2021,Bharti2021,Cohn2021,jamet2022,Baker2024,Yoshioka2025}, Subspace Expansion methods~\cite{McClean2017,Colless2018,Urbanek2020,Endo2021,Motta2021,Epperly2022,Motta2024,Umeano2025}, multi-state contraction~\cite{Parrish2019b,Huggins2020}, the Variational Quantum-Neural Hybrid Eignsolver~\cite{Zhang2022}, and configuration recovery~\cite{Moreno2025}.  Some of these algorithms have recently been shown to be competitive with classical computing algorithms for certain problems~\cite{Arute2019,anand2023,Chowdhury2024,Youngseok2026}.  We focus on a demonstration from IBM in which their quantum computers simulated short-time evolution for an Ising model with an order of 100 spins~\cite{Youngseok2026}.  Such calculations are not possible by brute force algorithms on a classical computer.  There may exist a classical algorithm that is capable of reproducing the simulation. Regardless, the ability to avoid constructing problem specific approximation methods grants the quantum calculation utility~\cite{Davis2023}.

Herein, we show that short-time evolution can be used to find the ground-state energy for an Ising model.  We utilize the CVQE algorithm with the Guided-Sampling Ansatz (GSA)~\cite{gunlycke2025,stenger2025}, which applies short-time evolution to the quantum computer and uses the samples to construct the ansatz.  We perform a demonstration for up to 63 qubits and analyze the quantum errors as a function of system size and coupling strength.  We apply our method for both homogeneous and random-coupling Ising models.  These types of calculations may lead to the discovery of new physical insights.   For example, the random-coupling model can be used to simulate spin-glasses, for which certain properties are still unknown~\cite{Bernaschi2024,Vincent2024}.

\section{Model}

\begin{figure}[h]
\vspace{2mm}
\begin{center}
\includegraphics[width=\columnwidth]{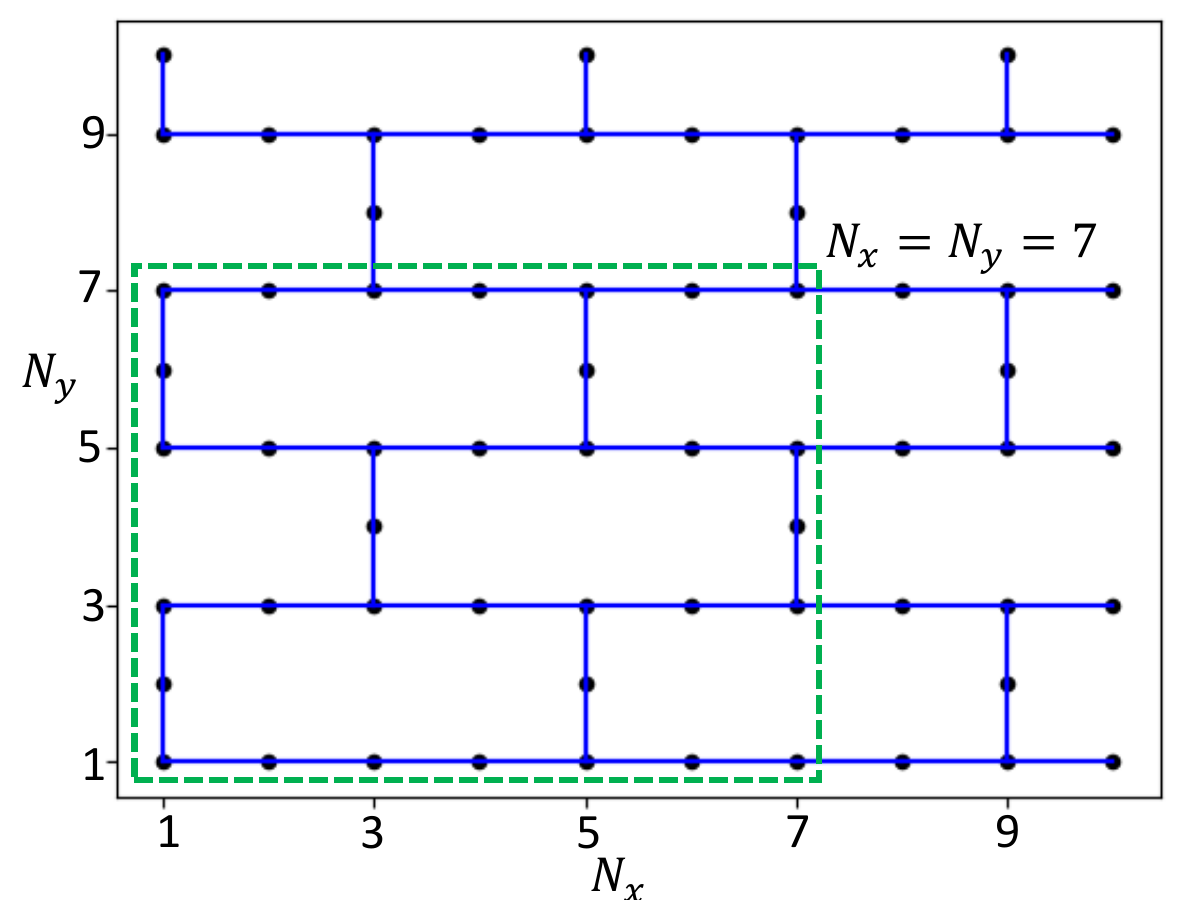}
\end{center}
\vspace{-2mm}
\caption{ Heavy-hex lattice.  The circles represent lattice sites and the lines represent connectivity.  We describe the size of a lattice by the number of layers in the x-direction $N_x$ and the number of layers in the y-direction $N_y$.  An example $Nx=7$, $N_y=7$ lattice is outlined by the dashed lines.  }
\label{F0}
\vspace{-3mm}
\end{figure}

We study the transverse-field Ising model
\begin{equation}
    \hat H = B \sum_i \hat{\sigma}_i^Z + \sum_{<i,j>} J_{ij} \hat{\sigma}_i^X \hat{\sigma}_j^X,
\end{equation}
where $\hat{\sigma}_i$ are Pauli operators, $B$ is the external field, $J_{ij}$ is the coupling between spins, and $\braket{i,j}$ denotes nearest-neighbor spins.  We assume a heavy-hex lattice to match the qubit architecture of IBM quantum computers.  We define the size of the lattice as shown in Fig.~\ref{F0}, where $N_x$($N_y$) is the number of lattice layers in the x(y)-direction.  We define $N_Q$ as the total number of spins.  We perform a simple mapping of the spin index to the qubit index.

\section{Guided-Sampling Ansatz}

For the GSA, the quantum computer is used to generate a guiding state.  In particular, we use short-time adiabatic state preparation.
To do this, we define the functions 
\begin{equation}
    \Omega_{ij}(t) = J_{ij} \frac{t}{T},
\end{equation}
and the time-dependent Hamiltonian
\begin{equation}
    \hat H(t) = B \sum_i \hat{\sigma}_i^Z + \sum_{<ij>} \Omega_{i,j}(t) \hat{\sigma}_i^X \hat{\sigma}_j^X.
\end{equation}
Let $\ket{\Psi(t)}$ be the instantaneous ground state of $H(t)$.  The initial Hamiltonian $\hat H(0)$ has the ground state $\ket{\Psi(0)} = \ket{\downarrow \downarrow \ldots}$, assuming $B$ is positive.  From the adiabatic theorem~\cite{duan2020}, for an energy gap of $\delta E$ between the ground and first excited state, if $T \gg 1/|\delta E|$ we arrive at the ground state of $\hat H(T)$ by applying the evolution operator
\begin{equation}
    \ket{\Psi(T)} = \hat U(T) \ket{\Psi(0)},
\end{equation}
where the evolution operator is
\begin{equation}
    \hat U(T) = \mathcal{T} e^{-i \int_0^T H(t) dt},
\end{equation}
and $\mathcal{T}$ is the time-ordering operation.  

In order to simulate evolution on a quantum computer, we have to perform a Trotter--Suzuki~\cite{trotter1959,Suzuki1976} decomposition
\begin{equation}
\begin{split}
        &\tilde U(T,\Delta t) \approx 
        \\
        & \prod_{k=0}^{N_T} \left( \prod_i e^{-iB\hat{\sigma}_i^Z \Delta t} \right) \left( \prod_{<ij>} e^{-i \Omega_{ij}(T-k \Delta t) \hat{\sigma}_i^X \hat{\sigma}_j^X \Delta t} \right)
\end{split}
\end{equation}
such that $T = N_T \Delta t$.  For $T \gg \Delta t$ the Trotter--Suzuki decomposition is a good approximation of the exact evolution $\hat U(T) = \lim_{\Delta t\rightarrow 0} \tilde U(T,\Delta t)$.  

However, even for $T \sim \Delta t$, we expect the operator $\tilde U(T,\Delta t)$ to guide the state towards the ground state of $\hat H(T)$.  In fact, the probability distribution should exponentially decay for states outside of an energy window $\Delta E \sim 1/\Delta t$.  See Appendix \ref{DSP} for more detail.    We use the probability distribution generated by $\tilde U(T,\Delta t)$ to construct a state space that contains an approximation of the ground state.  In the absence of errors, each shot of the quantum computer returns a basis state $\ket{\Phi_s}$ with probability
\begin{equation}
    P_{s} = | \bra{\Phi_s} \tilde{U}(\Delta t) \ket{\Psi(0)} |^2.
\end{equation}
  Let $\mathcal B = \{\ket{\Phi_s}\}_{s\in\mathcal S}$ be the set of basis states returned by the quantum computer after a number of shots $N_S$, where $\mathcal S$ is the set of indices for those states.  Furthermore, let 
\begin{equation}
    \mathcal B_{\lambda} = \{ \ket{\Phi_{s'}} |~\forall s,\forall \lambda':  ~ \ket{\Phi_s} \in \mathcal B, ~ 0 \leq \lambda' \leq \lambda, ~ \braket{\Phi_{s'}|\hat H^\lambda|\Phi_s} \neq 0 \}
\end{equation}
be the set of all basis states generated by applying $\hat H$ to each $\ket{\Phi_s}$ up to $\lambda$ times.  

We can obtain an approximation of the ground state by taking a projection of the Hamiltonian  
\begin{equation}
    \hat H^{(\lambda)}_{ss'} = \sum_{ss' \in \mathcal S_{\lambda}} \braket{\Phi_s|\hat H |\Phi_{s'}} \ket{\Phi_s}\!\bra{\Phi_{s'}},
\end{equation}
where $\mathcal S_\lambda$ is the index set for $\mathcal B_\lambda$.  We solve for the eigenvalues of $\hat H^{(\lambda)}$
\begin{equation}
    \hat H^{(\lambda)}\ket{\Psi^{(\lambda)}_n}  = E^{(\lambda)}_n \ket{\Psi^{(\lambda)}_n},
\end{equation}
using a classical computer.  As long as $|\mathcal B_n|$ is not exponentially large, solving for the effective ground state is efficient on a classical computer.  $\mathcal B$ is subexponential by necessity because we cannot take exponentially many shots.  Therefore, $|\mathcal B_\lambda|$ is subexponential as long as applying $\hat H$ does not exponentially increase the number of basis states.  This is the case for many physically relevant Hamiltonians including the Ising model.

This method is an implementation of the CVQE.  It has been shown that $\mathcal B_{1}$ is sufficient for producing accurate energy expectation values~\cite{stenger2025}.  Thus, below we focus exclusively on $\mathcal B_{1}$ .  

\section{Entropic Analysis}

If the ground-state energy is not already known, it can be difficult to judge the accuracy of the quantum algorithm.  One method for providing a measure of confidence is to compare the information gained from the quantum computer to the information used to construct the approximate ground state.  Ideally, we would like the quantum computer to provide much more information than we need to construct a good approximation of the ground state.  When this is not the case, errors are likely to occur.

One can quantify the amount of information obtained by the quantum computer by calculating the shot entropy
\begin{equation}
    S_{\mathcal B} = \sum_{s \in \mathcal S} -P_s \log_2 P_s.
\end{equation}
  The maximum possible shot entropy is 
\begin{equation}
    S_{\mathcal B}^{\text{max}} =  \log_2 |\mathcal B|.
\end{equation}
If the maximum shot entropy is obtained, then minimal information has been gained from the quantum measurements.  The difference in information between the maximum shot entropy and the obtained shot entropy represents the amount of information gained by the quantum measurements.  We define the information 
\begin{equation}
    I_{\mathcal B} = 2^{S_{\mathcal B}^{\text{max}}} - 2^{S_{\mathcal B}}.
\end{equation}

Turning to the amount of information used to construct the ground state, we define the distribution entropy
\begin{equation}
    S_{\Psi} = \sum_{s \in \mathcal S_1} -P^{\Psi}_s \log_2 P^{\Psi}_s,
\end{equation}
where $P^{\Psi}_s = |\braket{\Phi_s|\Psi^{(1)}_0}|^2$ is the probability that the effective ground state is in a given basis state.   
In terms of information, we define
\begin{equation}
    I_{\Psi} = 2^{ S_{\Psi} }.
\end{equation}

We want the information $I_{\mathcal B}$ obtained from the quantum measurements to be greater than the information $I_{\Psi}$ required to build the ground state.  However, these two measures are not directly comparable because $I_{\Psi}$ is calculated from $\mathcal B_1$ while $I_{\mathcal B}$ is calculated from $\mathcal B$.  Let $K$ be the number of terms in $\hat H$ that couple to each basis state, then $|\mathcal B_1| \approx K |\mathcal B|$.  Thus, we need $K$ times as much information from the quantum computer to construct a state in $\mathcal B_1$.  Using this factor, we can compare the two metrics of information by taking a ratio
\begin{equation}
    R = \log_2 \frac{I_{\mathcal B}}{K I_\Psi}
    \label{R}
\end{equation}
where the logarithm is introduced simply to make the final numbers more manageable.

\section{Results}

We us \textit{ibm\_torino} for all quantum computing applications.  \textit{ibm\_torino} is a Heron r1 quantum computer with qubits arranged in a heavy-hex lattice.  It has a median two qubit gate error of $2.5 \times 10^{-3}$ per controlled-Z gate and a median readout assignment error of $3.0\times10^{-2}$.  Further information can be found on the IBM website~\cite{ibm_torino}.  We transpile our circuits using the qiskit transpile function at optimization level 3 and with basis gates $rz$, $ry$, $rx$, and $rzz$.  We collect the results using the sampler version 2 qiskit primitive with 1000 shots.  No explicit error mitigation is performed.

\subsection{Homogeneous coupling}

We start with homogeneous couplings
\begin{equation}
    J_{ij} = J_0.
\end{equation}
    For $J_0 = 0$, the ground-state energy is trivially $E_0 = -B {N_Q}$ and the ground state is $\ket{\Psi_0} = \ket{\downarrow}^{\otimes N_Q}$.  In the limit $J_0/|B| \rightarrow -\infty$ the eigenstates are equal superpositions of the basis states with the ground state being specifically $\ket{\Psi_0} = \ket{+}^{\otimes N_q} + \ket{-}^{\otimes N_q}$, where $\ket{+} = (\ket{\uparrow} + \ket{\downarrow})/\sqrt{2}$ and the ground-state energy is $E_0 = J_0 {N_{\text{nbr}}}$, where $N_{\text{nbr}}$ is the number of nearest neighbors in the lattice.  In the regime $|J_0| \sim |B|$, we expect the ground-state eigenvector to be a sum of basis states with decaying amplitudes.  If we know the order of the amplitudes, we can build a subspace that includes only the highest amplitude basis states and solve the problem within that subspace.  It is difficult to predict the ordering of the amplitudes a-priori.  However, we can find the ordering using the quantum computer.  %We need only take a number of shots of the quantum computer proportional to the size of the subspace.        

\begin{figure}[h]
\vspace{2mm}
\begin{center}
\includegraphics[width=\columnwidth]{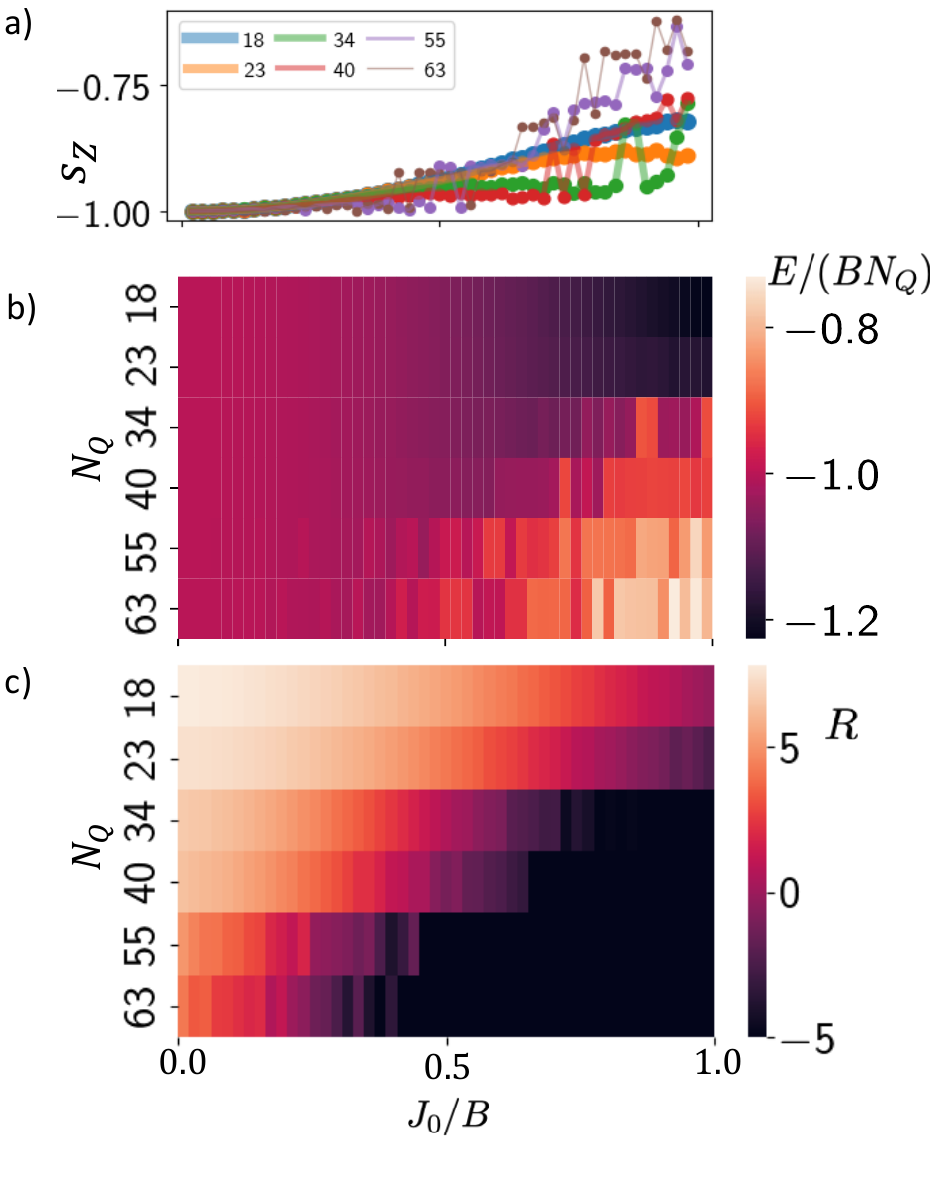}
\end{center}
\vspace{-2mm}
\caption{ Results for the homogeneous coupling model.  (a) the average spin as a function of the coupling strength $J_0$ for various lattice sizes.  The total number of qubits $N_{Q} = 18,23,34,40,55,63$ corresponds to lattice sizes of $N_x=N_y = 5,6,7,8,9,10$ respectively. (b) Effective ground-state energy $E^{1}_0$ as a function of the total number of qubits $N_Q$ and the coupling strength $J_0$.  (c) the information ratio $R$ as a function of the total number of qubits $N_Q$ and the coupling strength $J_0$. }
\label{F1}
\vspace{-3mm}
\end{figure}

Figure~\ref{F1} shows the results for the homogeneous couplings.  In Fig.~\ref{F1}a, we plot the average spin $s_Z = \sum_q \bra{\Psi^{1}_0}\hat{\sigma}^Z_q\ket{\Psi^{1}_0}/N_Q$ for various lattice sizes as a function of the coupling strength.  The effective ground-state energy is plotted in Fig.~\ref{F1}b.  Due to the homogeneity of the system, we expect both the energy and the total spin to follow smooth curves as a function of $J_0$.  Therefore, deviations from smooth curves represent errors in the method.  The errors on the quantum computer increase with both $N_Q$ and $|J_0|$.  Thus, the ground-state energy is poorly approximated for large $N_Q$ and large $|J_0|$.  $N_Q$ increases the error due to an increased number of qubits and, therefore, an increased number of faulty quantum logic gates.  $J_0$ does not change the number of quantum logic gates but it does change the duration of certain gates.  Furthermore, both $N_Q$ and $|J_0|$ increase the size of the effective subspace required to obtain a good approximation of the ground state.

While it is clear where the ground-state energy approximation deviates from the expected smooth curve in this homogeneous case, there is no such expectoration for comparison in the general case.  Therefore, it is important to establish some measure of confidence based solely on the results of the calculation.  In Fig.~\ref{F1}c, we plot $R$, from Eq.~\eqref{R}, as a function of both $N_Q$ and $J_0/B$.  By comparing Fig.~\ref{F1}b and Fig.~\ref{F1}c we find $R<0$ is a good indication that the quantum algorithm has failed to produce a good approximation of the ground-state energy.  This is expected as $R = 0$ indicates that we have used approximately the same amount of information to build the ground state as we obtained from the quantum computer.

\subsection{Random coupling}

We define the random-coupling Ising model using uniformly distributed couplings
\begin{equation}
    J_{ij} = \text{Rand}(-\Delta_J/2,\Delta_J/2)
\end{equation}
where Rand$(A,B)$ is a real number randomly generated from a uniform distribution between $A$ and $B$.  This model describes a spin glass beyond some critical point $\Delta_J > J_c$, typically $J_c\sim |B|$.  In the limit $\Delta_J/|B| \rightarrow \infty$ the eigenstates are equal distribution of the basis states similar to the homogeneous case.  Unlike the homogeneous case, it can be difficult to predict the phase for each amplitude in the ground-state eigenvector even in the infinite coupling limit.

\begin{figure}[h]
\vspace{2mm}
\begin{center}
\includegraphics[width=\columnwidth]{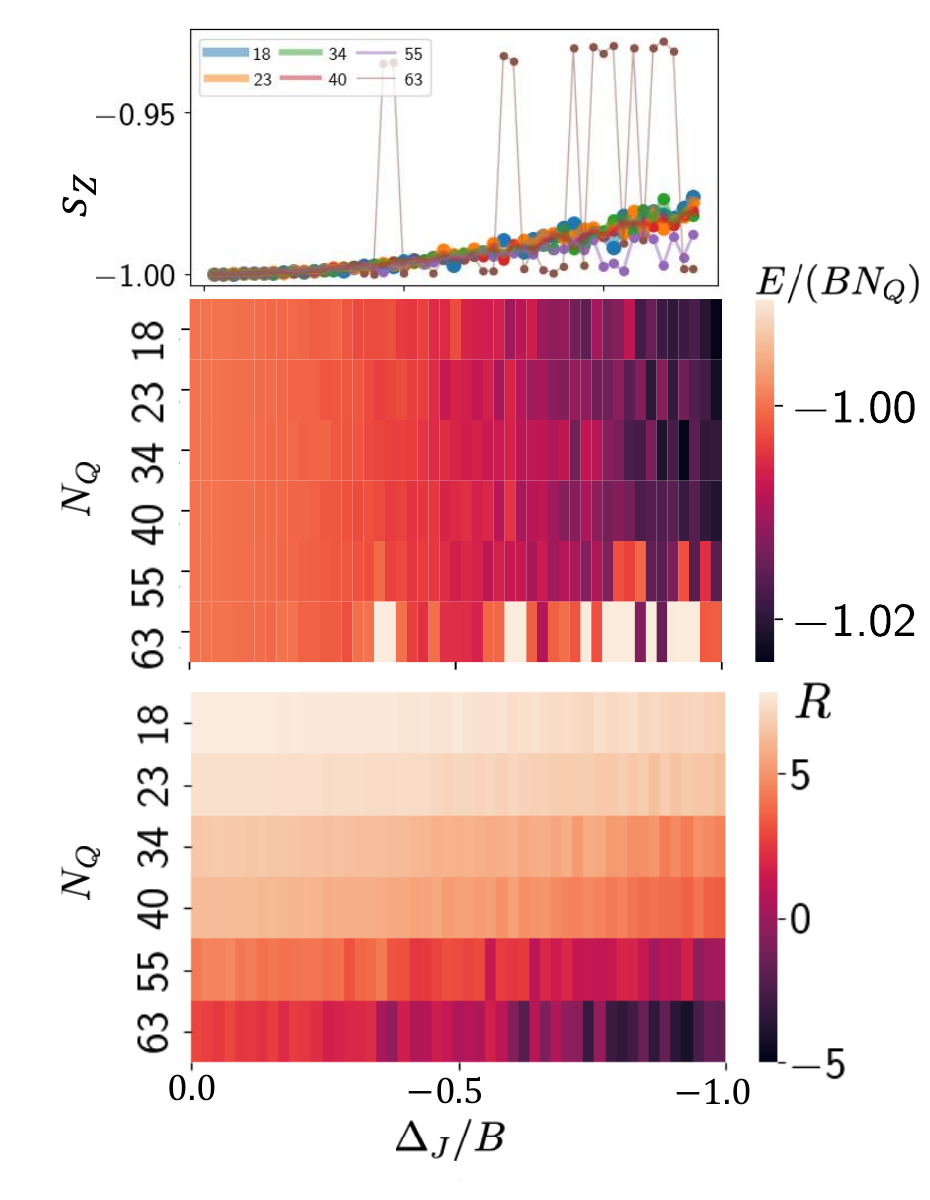}
\end{center}
\vspace{-2mm}
\caption{ Results from the random-coupling model.  (a) the average spin as a function of the coupling-strength bound $\Delta_J$ for various lattice sizes.  The total number of qubits $N_{Q} = 18,23,34,40,55,63$ corresponds to lattice sizes of $N_x=N_y = 5,6,7,8,9,10$ respectively. (b) Effective ground-state energy $E^{1}_0$ as a function of the total number of qubits $N_Q$ and the coupling-strength bound $\Delta J$.  (c) the information ratio as function of the total number of qubits $N_Q$ and the coupling-strength bound $\Delta J$. }
\label{F2}
\vspace{-3mm}
\end{figure}

Another difference to the homogeneous case is that the ground-state energy does not form a smooth curve as a function of $\Delta_J$.  Figure~\ref{F2} shows the results of our calculation for the random-coupling model.  Figure~\ref{F2}a shows the average spin $s_Z = \sum_q \bra{\Psi^{1}_0}\hat{\sigma}^Z_q\ket{\Psi^{1}_0}/N_Q$  as a function of the coupling strength range $\Delta_J$.  Figure~\ref{F2}b shows the effective-ground-state energy as a function of $\Delta_J$ and system size $N_Q$. While there is an average downward sloping trend to the energy as $\Delta_J$ increases, the value of $E_0$ deviates around the trend line.  It is difficult to distinguish between deviations that follow the value of the true ground-state energy  and those caused by errors from the quantum computer.  To increase our confidence in the results, we turn to Fig.~\ref{F2}c where we plot $R$ as a function of $\Delta_J$ and $N_Q$.  We find $R>0$ except for a small number of points near the bottom right corner of the plot.  We also see that the value of $R$ tracks the largest deviations in the energy, which likely correspond to quantum errors.

\subsection{Subspace analysis}

\begin{figure*}[t]
\vspace{2mm}
\begin{center}
\includegraphics[width=1.9\columnwidth]{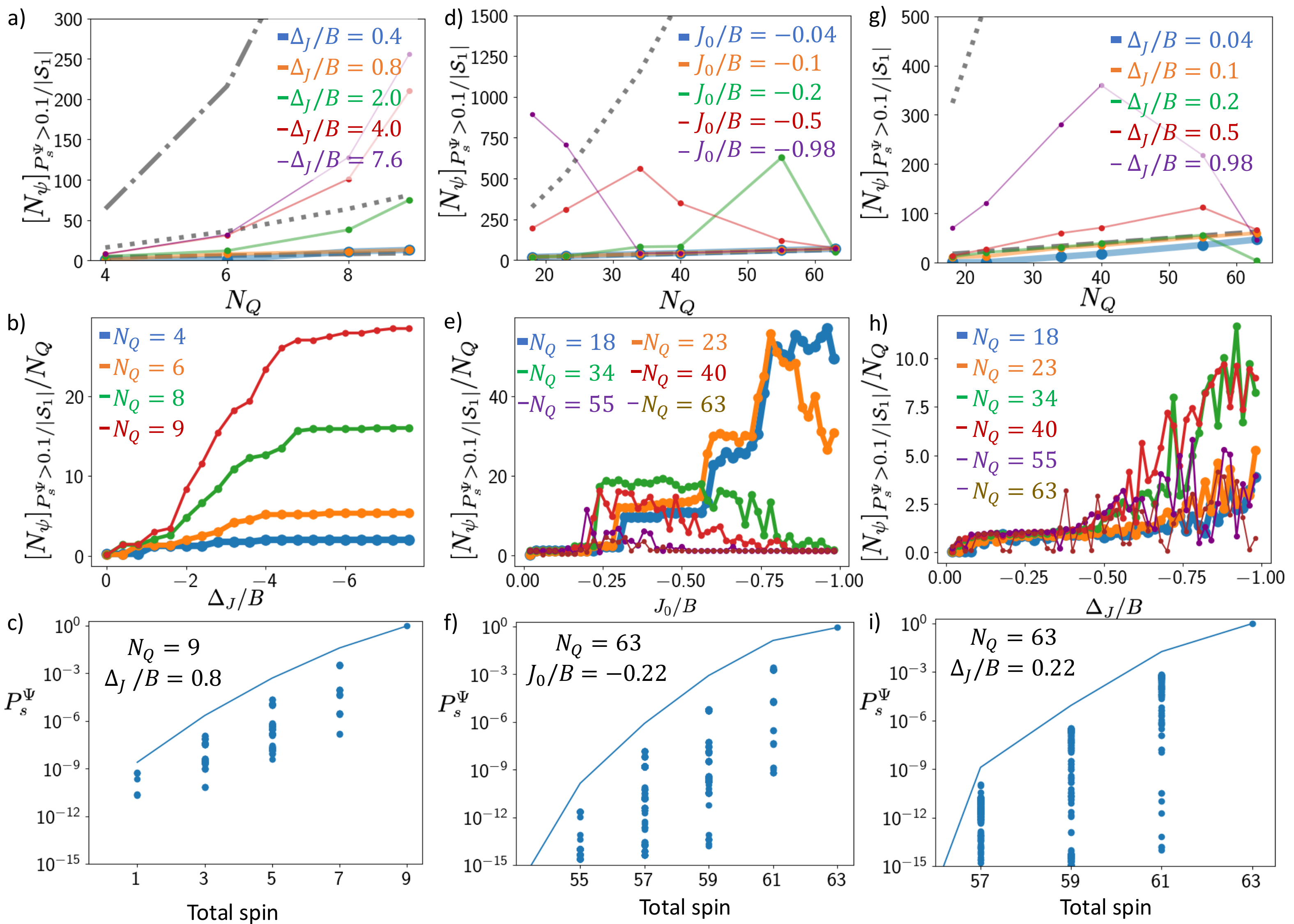}
\end{center}
\vspace{-2mm}
\caption{ Basis state structure of the ground state.  (a),(b),(c) are for a small square lattice model exactly solved and averaged over 1000 different random couplings.  (d),(e),(f) are for the homogeneous model.  (g),(h),(i) are for the random-coupling model.  (a),(d),(g) show the number of basis states $N_\psi$ that have a probability in the ground state greater than $P^{\Psi}_s > 0.1/|\mathcal B_1|$ as a function of $N_Q$ for select values of the coupling $J_0$ or $\Delta_J$.  The dashed, dotted, and dash-dot lines represent linear $N_Q$, square $N_Q^2$, and cubic $N_Q^3$ boundaries respectively. (b),(e),(h) show the number of basis states $N_\psi$ that have a probability in the ground state greater than $P^{\Psi}_s > 0.1/|\mathcal B_1|$ as a function of the coupling $J_0$ or $\Delta_J$ for select values of $N_Q$.  The y-axis is scaled by $1/N_Q$.  (c),(f),(i) show the probability of each basis state as a function of total spin for specific values of $N_Q$ and $J_0$ or $\Delta_J$.  Each dot represents a basis state.  The curves represents the total probability of all basis states with a given total spin.    }
\label{F5}
\vspace{-3mm}
\end{figure*}

Figure~\ref{F5} shows the basis state structure for the ground state under several different scenarios.  The panels on the left hand side (Fig.~\ref{F5}a,b,c) show data from an exactly solved square lattice model where we average over the results of 1000 different random couplings.  Fig.~\ref{F5}a shows plots of the number of basis states with a probability above $P^{\Psi}_s > 0.1/|\mathcal B_1|$ as a function of the number of spins $N_Q$.  For $\Delta_J/B < 1$ we see that the number of significant basis states grows nearly linearly with the size of the system.  As we increase $\Delta_J/B > 1$ we see that the number of significant basis states also increases.  However, the dependence on $N_Q$ does not surpass cubic even for the largest coupling values we calculated $\Delta_J/B = 7.6$.  Fig.~\ref{F5}b  also shows the total number of basis states with $P^{\Psi}_s > 0.1/|\mathcal B_1|$ but as a function of the coupling $\Delta_J$ and divided by $N_Q$.  We see that the number of basis states with probability above the cutoff increases incrementally.  This is due to the small system size, as the probability distribution is discrete.  This is seen clearly in Fig.~\ref{F5}c where we plot the probability of each basis state sorted into bins of total spin.  Notice that there is a single basis state with total spin equal to 9 with the highest probability.  For each pair of spins that is flipped, the total probability of all of the basis states with that total spin decreases exponentially.  

The panels in the middle column (Fig.~\ref{F5}d,e,f) show data for the homogeneous heavy-hex lattice solved using the CVQE algorithm.  Figure~\ref{F5}d plots the number of basis states with a probability above $P^{\Psi}_s > 0.1/|\mathcal B_1|$ as a function of the number of spins $N_Q$.  For small $|J_0|$ the number of significant basis states grows nearly linearly with system size.  For the largest $|J_0|$ values we studied, the number of significant basis states never grows faster than $N_Q^2$.  Often, the number of significant basis states will decrease.  This is likely because it is more difficult for the quantum computer to find the significant basis states for larger systems.  Figure~\ref{F5}e shows the total number of basis states with $P^{\Psi}_s > 0.1/|\mathcal B_1|$ divided by $N_Q$ as a function of the coupling $J_0$.  Once again, we see that the number of significant basis states grows incrementally.  This is because many of the basis states have identical probability.  Thus, even though there is a large number of basis states, the probability distribution is still discrete.  This is seen in Fig.~\ref{F5}f where we plot the probability of each basis state sorted into bins of total spin.  As before, the total probabilities in each bin decrease exponentially with the number of flipped spins.  Note, however, that not all states in a given bin have larger probability than all states in lower bins.  This is where the quantum computer provides guidance by sorting  basis states according to their relevance to the ground state.  As $J_0/B$ increases, the probability-vs-total-spin curve flattens, making it increasingly more difficult to choose the relevant basis states classically. 

The rightmost panels (Fig.~\ref{F5}g,h,i) show data for the random-coupling model solved using the CVQE algorithm.  Figure~\ref{F5}g shows the number of basis states with a probability greater than $P^{\Psi}_s > 0.1/|\mathcal B_1|$ as a function of the number of spins $N_Q$.  As in the previous cases, small couplings result in nearly linear dependence on system size while all coupling with $\Delta_j/B < 1$ increase slower than $N_Q^2$.  Figure~\ref{F5}h shows the total number of basis states with $P^{\Psi}_s > 0.1/|\mathcal B_1|$ divided by $N_Q$ as a function of the coupling $\Delta_J$.  Unlike in the previous cases, the dependence on coupling does not appear to increase in discrete steps.  This is because the randomness of the coupling and the large number of basis states make the probability distribution effectively continuous.  This is seen in Fig.~\ref{F5}i where we plot the probability of each basis state sorted into bins of total spin.  As in the previous two cases, the total probability in each bin decreases exponentially with the number of flipped spins.  Unlike the previous cases, the distribution of probabilities within a bin is nearly continuous.  

Figure~\ref{F5} strongly suggests that the number of basis states required to have a good approximation of the ground state does not grow exponentially with system size for Ising models in the regime $|J_{ij}|\sim|B|$.  This makes Ising models especially suitable for exploration using the CVQE method.  Given the simplicity of the mapping between the Ising model and the quantum gates used to control the quantum computer, these types of calculations may be among the first quantum computing calculations to reveal new physical insights about nature.

\section{Conclusion}

The transverse field Ising model with random couplings is a difficult problem for classical computers in the regime of $|\Delta_J| \sim |B|$ for large system sizes.  However, we expect that this regime does not pose any fundamental problems for quantum computers. We have performed quantum calculations for the ground-state energy and average spin for systems of up to $N_Q = 63$ qubits.  We propose a measure of the accuracy of our method based on a comparison of the information content provided by the quantum computer and the information used to construct the effective ground state.  We are able to approach the $|\Delta_J| \sim |B|$ limit, however, errors in the quantum computer prevent us from finding accurate results at $|\Delta_J| = |B|$ for the largest systems.  

The data we have collected suggest that current quantum computers can approach the $|\Delta_J| \sim |B|$ limit for large system sizes without significant increases in the information requirements for the ground state.  As quantum computers become more accurate, we expect that we will be able to extend the quantum computer deeper into the large-$N_Q$-large-$\Delta_J$ limit.  

 \section{Acknowledgment}

This work has been supported by the Office of Naval Research (ONR) through the U.S. Naval Research Laboratory (NRL).  We acknowledge QC resources from IBM through a collaboration with the Air Force Research Laboratory (AFRL).

\appendix

\section{Diabatic State Preparation}
\label{DSP}

Ideally we would perform adiabatic state preparation on the quantum computer to find the true ground state.  However, due to the short decoherence times of modern-day quantum computers, we are unable to change the Hamiltonian slowly  enough to guarantee adiabatic evolution.  Thus, we instead perform short-time ($T\sim \Delta t$) state preparation.  We now show that short-time state preparation can find the appropriate subspace within a controllable accuracy. 

We define the instantaneous eigenstates $\ket{\Psi_n(t)}$ and eigenvalues $E_n(t)$
\begin{equation}
    \hat H(t) \ket{\Psi_n(t)} = E_n(t)\ket{\Psi_n(t)},
    \label{A1}
\end{equation}
of a dynamic Hamiltonian $H(t)$ such that $E_m(t) \geq E_n(t)$ when $m > n$.  We also define the dynamic state vector 
\begin{equation}
    i \frac{d}{dt}\ket{\tilde \Psi(t)} = \hat H(t) \ket{\tilde \Psi(t)},
    \label{A2}
\end{equation}
where $\ket{\tilde \Psi(0)} = \ket{\Psi_0(0)}$.
The dynamic state vector can always be written as a linear combination of the instantaneous eigenstates
\begin{equation}
    \ket{\tilde \Psi(t)} = \sum_n c_n(t)\ket{\Psi_n(t)}
\end{equation}
where $c_n(t)$ are complex coefficients.  Our goal is to understand when $|c_0(t)| >> 0$.    

We expand Eq.~\eqref{A2} in terms of instantaneous eigenstates
\begin{equation}
    i \frac{d}{dt} \sum_n c_n(t) \ket{\Psi_n(t)} =  \sum_n c_n(t) E_n(t) \ket{\Psi_n(t)}.
\end{equation}
using the product rule and operating from the left with $\bra{\Psi_m(t)}$ we have
\begin{equation}
    i \dot{c}_m(t) + i \sum_n \braket{\Psi_m(t)|\dot{\Psi}_n(t)}c_n(t) = E_m(t)c_m(t).
    \label{A5}
\end{equation}
We need to evaluate the bracket $\braket{\Psi_m(t)|\dot{\Psi}_n(t)}$.  For this purpose, let us take a time derivative of Eq.~\eqref{A1}
\begin{equation}
    \dot{H}(t) \ket{\Psi_n(t)} + \hat H(t)\ket{\dot{\Psi}_n(t)} = \dot{E}_n(t) \ket{\Psi_n(t)} + E_n(t) \ket{\dot{\Psi}_n(t)}
\end{equation}
Operating from the left with $\bra{\Psi_m(t)}$ and rearranging we have
\begin{equation}
    \braket{\Psi_m(t)|\dot{\Psi}_n(t)} = \frac{\braket{\Psi_m(t)|\dot{H}(t)|\Psi_n(t)}}{E_n(t)-E_m(t)}
\end{equation}
for $m \neq n$. 
Using this result in Eq.~\eqref{A5} and rearranging, we get an equation for the amplitudes
\begin{equation}
\begin{split}
    \dot{c}_m(t) =&  -\big[iE_m(t) +\braket{\Psi_m(t)|\dot{\Psi}_m(t)}\big]c_m(t) 
    \\
    & -\sum_{n\neq m} \frac{\braket{\Psi_m(t)|\dot{H}(t)|\Psi_n(t)}}{E_m(t)-E_n(t)} c_n(t).
\end{split}
\label{A8}
\end{equation}
In the adiabatic approximation, we assume that $E_{n\neq 0}(t) - E_0 \gg |\dot{H}(t)|$ for all $t$.  In this case, we can ignore the second term and we have that $(c_{n\neq 0}(0) = 0) \implies (c_{n\neq 0}(t) = 0)$.  For fast evolution, we may not have $E_{n\neq 0}(t) - E_0 \gg |\dot{H}(t)|$, however, we may still have $E_{n > m}(t) - E_0 > |\dot{H}(t)|$ for some $m$.  Due to the form of Eq.~\ref{A8} the amplitude for states $\ket{\Psi_{n>m}(t)}$ will exponentially decay away from $m$.  Thus, $|c_0(t)|$ is inversely proportional to the number of states $\ket{\Psi_{n<m}(t)}$ with energies $E_n(t) < E_m(t)$.   In this way, highly diabatic time evolution can still find a subexponential vector space containing the ground state as long as there is a subexponential number of energy levels with $E_n(t) < E_m(t)$ for all $t$ during the evolution.     

\bibliography{ref}

\end{document}